\documentclass[onecolumn,english,aps,prb,showpacs,floatfix,superscriptaddress, 11pt, notitlepage]{revtex4-1}
\usepackage[T1]{fontenc}
\usepackage[latin9]{inputenc}
\usepackage[english]{babel}
\usepackage{amsmath}
\usepackage{amssymb}
\usepackage{graphicx}
\usepackage{float}
\usepackage{titlesec}
\usepackage{soul}
\usepackage{siunitx}

\setcitestyle{super,open={},close={}}

\DeclareRobustCommand*{\citen}[1]{%
  \begingroup
    \setcitestyle{numbers,open=[,close=]}%
    \cite{#1}%
  \endgroup   
}

\makeatletter
\@ifundefined{textcolor}{}
{%
 \definecolor{BLACK}{gray}{0}
 \definecolor{WHITE}{gray}{1}
 \definecolor{RED}{rgb}{1,0,0}
 \definecolor{GREEN}{rgb}{0,1,0}
 \definecolor{BLUE}{rgb}{0,0,1}
 \definecolor{CYAN}{cmyk}{1,0,0,0}
 \definecolor{MAGENTA}{cmyk}{0,1,0,0}
 \definecolor{YELLOW}{cmyk}{0,0,1,0}
}

\newcommand*{\balancecolsandclearpage}{%
  \close@column@grid
  \clearpage
  \twocolumngrid
}

\@ifundefined{definecolor}
 {\usepackage{color}}{}

\makeatother

\usepackage{babel}
\begin{document}
\flushbottom

\titlespacing*{\section}
{500pt}{10pt}{0pt}
\titlespacing*{\subsection}
{0pt}{5.5ex plus 1ex minus .2ex}{4.3ex plus .2ex}

\title{Topological order and equilibrium in a condensate of exciton-polaritons}

\author{Davide~Caputo}
\affiliation{CNR NANOTEC---Institute of Nanotechnology, Via Monteroni, 73100 Lecce, Italy}
\affiliation{University of Salento, Via Arnesano, 73100 Lecce, Italy}

\author{Dario~Ballarini}
\affiliation{CNR NANOTEC---Institute of Nanotechnology, Via Monteroni, 73100 Lecce, Italy}

\author{Galbadrakh~Dagvadorj}
\affiliation{Department of Physics, University of Warwick, Coventry CV4 7AL, United Kingdom}

\author{Carlos~S\'anchez~Mu\~noz}
\affiliation{Departamento de F\'{i}sica Te\'orica de la Materia Condensada, Universidad Aut\'onoma de Madrid, 28049 Madrid, Spain}

\author{Milena~De~Giorgi}
\affiliation{CNR NANOTEC---Institute of Nanotechnology, Via Monteroni, 73100 Lecce, Italy}

\author{Lorenzo~Dominici}
\affiliation{CNR NANOTEC---Institute of Nanotechnology, Via Monteroni, 73100 Lecce, Italy}

\author{Kenneth~West}
\affiliation{PRISM, Princeton Institute for the Science and Technology of Materials, Princeton Unviversity, Princeton, NJ 08540}

\author{Loren~N.~Pfeiffer}
\affiliation{PRISM, Princeton Institute for the Science and Technology of Materials, Princeton Unviversity, Princeton, NJ 08540}

\author{Giuseppe~Gigli}
\affiliation{CNR NANOTEC---Institute of Nanotechnology, Via Monteroni, 73100 Lecce, Italy}
\affiliation{University of Salento, Via Arnesano, 73100 Lecce, Italy}

\author{Fabrice~P.~Laussy}
\affiliation{Russian Quantum Center, Novaya 100, 143025 Skolkovo, Moscow Region,
Russia}

\author{Marzena~H.~Szyma\'nska}
\affiliation{Department of Physics and Astronomy, University College London,
Gower Street, London WC1E 6BT, United Kingdom}

\author{Daniele~Sanvitto}
\affiliation{CNR NANOTEC---Institute of Nanotechnology, Via Monteroni, 73100 Lecce, Italy}


\maketitle

\textbf{
We report the observation of the Berezinskii--Kosterlitz--Thouless transition for a 2D gas of exciton-polaritons, and through the joint measurement of the first-order coherence both in space and time we bring compelling evidence of a thermodynamic equilibrium phase transition in an otherwise open driven/dissipative system. This is made possible thanks to long polariton lifetimes in high-quality samples with small disorder and in a reservoir-free region far away from the excitation spot, that allow topological ordering to prevail. The observed quasi-ordered phase, characteristic for an equilibrium 2D bosonic gas, with a decay of coherence in both spatial and temporal domains with the same algebraic exponent, is reproduced with numerical solutions of stochastic dynamics, proving that the mechanism of pairing of the topological defects (vortices) is responsible for the transition to the algebraic order. Finally, measurements in the weak-coupling regime confirm that polariton condensates are fundamentally different from photon lasers and constitute genuine quantum degenerate macroscopic states.
}


%


Collective phenomena which involve the emergence of an ordered phase in many-body systems have a tremendous relevance in almost all fields of knowledge, spanning from physics to biology and social dynamics \cite{Stanley1999, Castellano2009}. While the physical mechanisms can be very different depending on the system considered, statistical mechanics aims at providing universal descriptions of phase transitions on the basis of few and general parameters, the most important being dimensionality and symmetry \cite{Onsager1944, Landau1980, Halperin1977}.  The spontaneous symmetry breaking of Bose--Einstein condensates (BEC) below a critical temperature $T>0$ is a remarkable example of such a transition, with the emergence of an extended coherence giving rise to a long range order (LRO) \cite{Stringari2016, Esslinger2008,Braun2015}. Notably, in infinite systems with dimensionality \(d\le 2\), true LRO cannot be established at any finite temperature \cite{Mermin1966}. This is fundamentally due to the presence of low-energy long-wavelength thermal fluctuations (i.e. Goldstone modes) that prevail in \(d\le 2\) geometries. 

However, if we accept a lower degree of order, characterised by an algebraic decay of coherence, it is still possible to make a clear distinction between such a quasi-long-range-ordered (QLRO) and a disordered phase in which the coherence is lost in a much faster, exponential way. Such transition, in two dimensions (2D) and at a critical temperature $T>0$, is explained in the Berezinskii--Kosterlitz--Thouless theory (BKT) by the proliferation of vortices---the fundamental topological defects---of opposite signs \cite{Minnhagen1987}. 
This theory is well established for 2D ensembles of cold atoms in thermodynamic equilibrium, where the transition is linked to the appearance of a linear relationship between the energy and the wavevector of the excitations in the quasi-ordered state \cite{Ozeri2002}. The joint observation of spatial and temporal decay of coherence has never been observed in atomic systems, mainly because of technical difficulties in measuring long-time correlations. These are important observables to bring together because an algebraic decay, with the same exponent $\alpha$, for both the temporal and spatial correlations of the condensed state, implies a linear dispersion for the elementary excitations\cite{Nelson1977,Szymanska2006,PhysRevB.75.195331}.
\begin{figure*}[htbp]
\centering
	\includegraphics[width=1\textwidth]{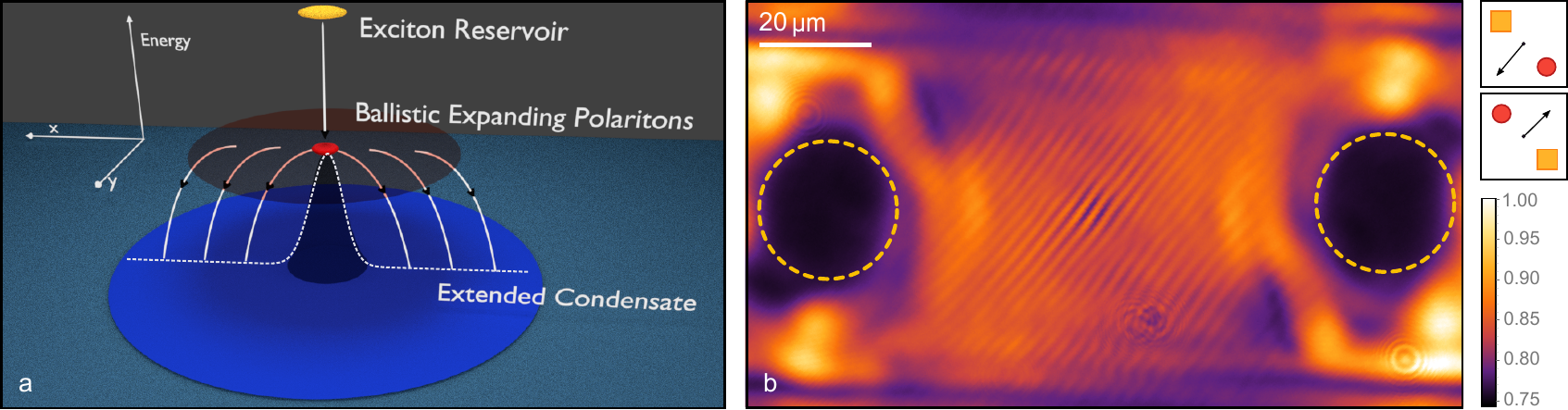}
	\caption{ \textbf{Sketch of the pumping mechanism and interferogram}.
          \textbf{a,} Spatial cross section (x,y) at different energies, where the
          vertical axes represents increasing energies. The carriers,
          injected by the pumping laser, relax quickly into excitonic
          states (yellow area) spatially confined within the pumping
          spot region. Efficient scattering from the exciton
          reservoir into polariton states results in a region of high
          polariton density (red area) which expands radially outwards
          and relax into lower energy states. Above a threshold power,
          an extended 2D polariton condensate (blue area) is formed
          outside of the pumped region.  \textbf{b,} Interference
          pattern of the extended 2D polariton condensate, obtained rotating one arm of the interferometer as the symbols on the right. The circles in
          yellow, dashed line indicate the blue-shifted region
          (emission is filtered around k=0) corresponding to the position of the laser spot.}
\label{fig1}
\end{figure*}

On the other hand, semiconductor systems such as microcavity polaritons (dressed photons with sizeable interactions mediated by the excitonic component) since the report of their condensation\cite{Richard2005b} appear to be ideal platforms to extend the investigation of many-body physics to the more general scenario of phase transitions in driven-dissipative systems \cite{Carusotto2013}. However, establishing if the transition can actually be governed by the same BKT process as for equilibrium system has proven to be challenging from both the theoretical\cite{PhysRevX.5.011017,Altman2016,Galaa2015} and experimental perspective. Indeed, the dynamics of phase fluctuations is strongly modified by pumping and dissipation, and the direct measurement of their dispersion by photoluminescence and four-wave-mixing experiments is limited by the short polariton lifetime, by the pumping-induced noise and by the low resolution close to condensate energy\cite{Utsunomiya2008, Kohnle2011, Kohnle2012}.
Moreover, the algebraic decay of coherence has been experimentally demonstrated only in spatial correlations, while only exponential or Gaussian decays of temporal coherence, which are not compatible with a BKT transition, have been reported until now\cite{yamamoto,Krizhanovskii2006, Love2008, PhysRevX.6.011026}. In particular, the lack of a power law decay of temporal correlations is a strong argument against a true BKT transition, as will be demonstrated later on in a straightforward counter-example of a strongly out-of-equilibrium polariton system.
For this reason, it has been a matter of constant debate in the field what is the nature of the various polariton phases, what are the observables that allow to determine a QLRO, if any, and how they compare with equilibrium 2D condensates and with lasers \cite{Dihel2008,Keeling2010, Kirton2013, Deng2003, Butov2007,Klaers2010}. 
Recently, thanks to a new generation of samples with record polariton lifetimes, a thermalization across the condensation threshold has been reported via constrained fitting to Bose--Einstein distribution, suggesting a weaker effect of dissipation in these systems \cite{Sun2016}. However, to unravel the mechanisms that drive the transition, and characterize its departure from the equlibrium condition, it is crucial to measure the correlations between distant points in space and time as we move from the disordered to the quasi-ordered regime\cite{Szymanska2006,PhysRevB.75.195331,Ciocchetta2013, keeling2016}.
So far, all attempts in this direction have been thwarted, not only by the polariton lifetime being much shorter than the thermalization time, but also by sample inhomogeneities \cite{Krizhanovskii2009, Sanvitto2005} and/or small extents of the condensate. The latter, in particular, limits the power-law decay of the spatial coherence to the small spatial extension of the exciton reservoir set by the excitation spot \cite{Deng2007, PhysRevB.90.205430, yamamoto}, which could result in an effective trapping mechanism \cite{Hadzibabic2006} and finite size effects \cite{Keeling2010}.

In this work, using a high quality sample (in terms of long lifetimes and spatial homogeneity) to form and control a reservoir-free condensate of polaritons over a largely extending spatial region, we make the first observation (in any system) of the transition to a QLRO phase both in spatial and in temporal domains. Remarkably, the convergence of spatial and temporal decay of coherence allows us to identify the connection with the classic equilibrium BKT scenario, in which for systems with linear spectrum the exponents take exactly the same value $\alpha\leq 1/4$.\cite{PhysRevB.75.195331} 
Stochastic simulations tuned to the experimental conditions, that reproduce the experimental observations in both space and time, further allow us to track vortices in each realisation of the condensate, confirming the topological origin of the transition. 
All these results settle the BKT nature of the 2D phase-transition for driven/dissipative polaritons in high quality samples, providing the equilibrium limit that allows to develop the paradigm of non-equilibrium regimes. Finally, this proves that a quasi-ordered polariton state is fundamentally different from a laser---excitons in a cavity under weak coupling conditions---for which a power-law decay of the first order coherence is observed only in space but not in time correlations, also putting to rest the long-time debate whether polaritons above threshold are distinct from a laser, and can indeed be called a condensate\cite{Butov2012, Deveaud-Pledran2012}.

\section*{Results} 
\begin{figure}[htbp]
\centering
	\includegraphics[width=0.48\textwidth]{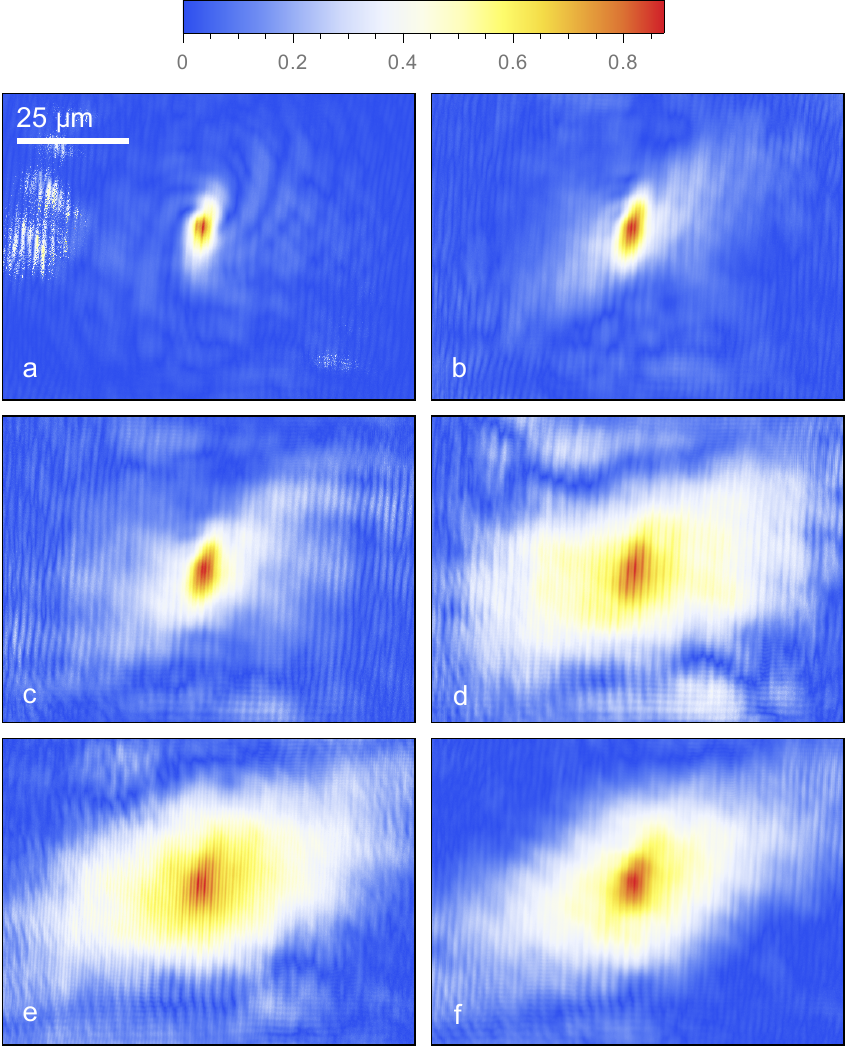}
	\caption{\textbf{Two dimensional first order correlation
            function} $g^{(1)}(\mathbf{r})$ at different
          densities. From the top left corner, d=\SI{0.015}{pol/\micro\meter\squared}, d=\SI{0.1}{pol/\micro\meter\squared},  
	d=\SI{0.15}{pol/\micro\meter\squared}, d=\SI{0.4}{pol/\micro\meter\squared}, d=\SI{0.9}{pol/\micro\meter\squared}, d=\SI{1.2}{pol/\micro\meter\squared} in \textbf{a, b, c, d, e, f}, respectively.
}
\label{fig2}
\end{figure}
\begin{figure*}
\centering
	\includegraphics[width=1\textwidth]{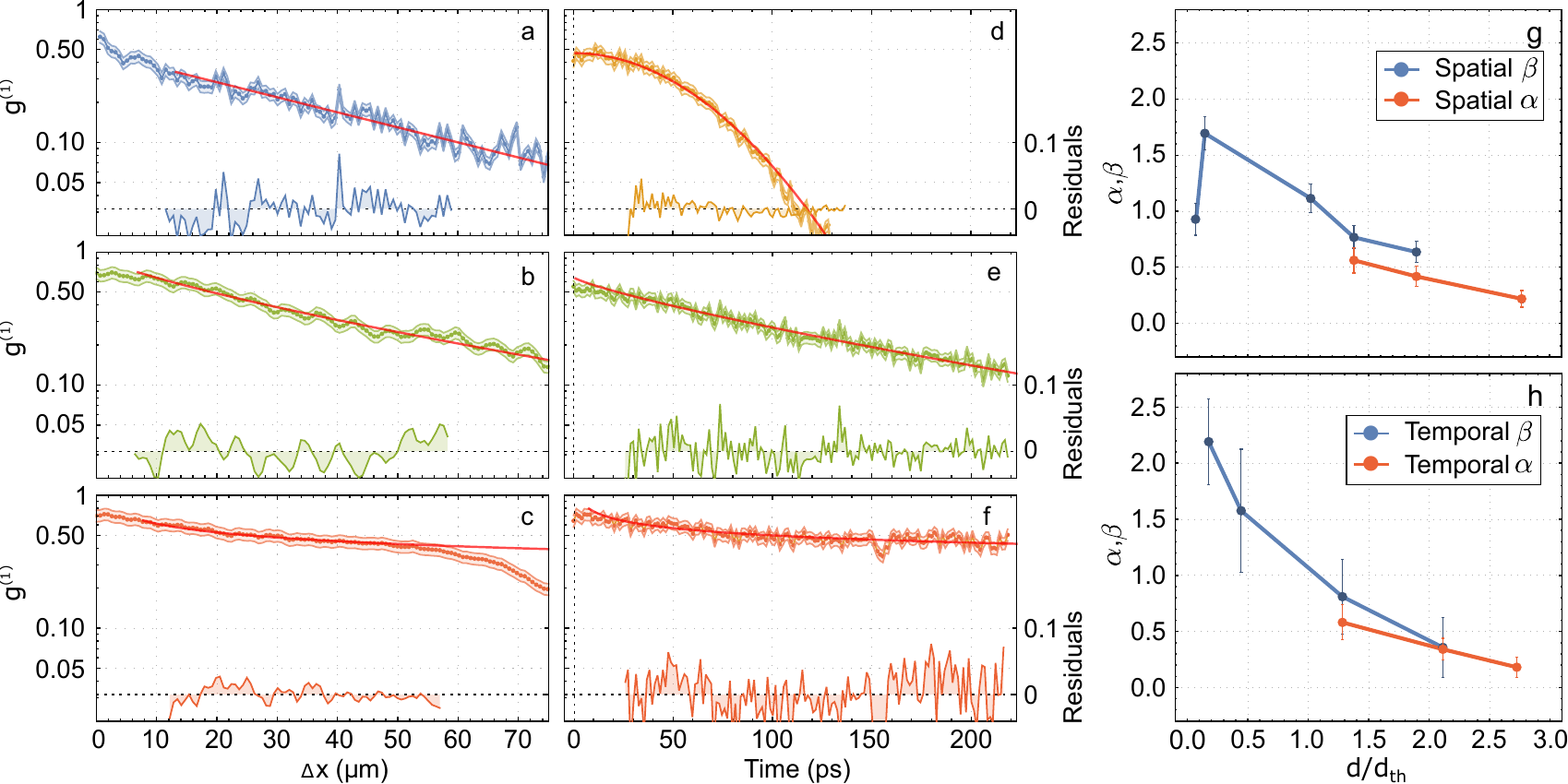}	
	\caption{
    \textbf{Coherence decay and BKT phase transition.}
    \textbf{a, b, c,} Details for $\Delta x>0$ of the exponential (blue data),
    stretched 
    exponential (green data), power law (red data) spatial 
    decay of $g^{(1)}(x,-x)$ and corresponding fitting residuals.
    \textbf{d, e, f,}  $g^{(1)}(t,t+\Delta t)$ for $\Delta t>0$ of the 
    Gaussian (yellow data),
    stretched 
    exponential (green data), power law (red data) temporal 
    decay and corresponding fitting residuals.
     \textbf{g,} Blue line: $\beta$ exponent from the stretched exponential
    fitting of $g^{(1)}(x,-x)$ as a function of 
    polariton densities.
     Red line: $\alpha$ exponent from the power law fitting of $g^{(1)}(x,-x)$ 
     in the above threshold density range.
    \textbf{h,} Blue line: $\beta$ exponent from the stretched exponential 
    fitting of $g^{(1)}(t,t+\Delta t)$ as a function of 
    polariton densities. Red line: $\alpha$ exponent from the power law  
    fitting of $g^{(1)}(t,t+\Delta t)$ in the above threshold density range.
    Error bars from fitting parameters errors.
}
 \label{fig3}
\end{figure*}
The sample is excited non-resonantly (Methods), leading to the
formation of the exciton reservoir (yellow region in Fig.~1a) which is localised within the pumping spot area due to the low exciton
mobility. In turn, the repulsive interactions between excitons induce
the energy blueshift of the polariton resonance at the center of the
pumping spot (the dashed-white line in Fig.~\ref{fig1}a is the energy blueshift in the $x$ direction, corresponding to the Gaussian intensity profile of the exciting beam). Polaritons, which are formed beneath the exciton reservoir
through energy relaxation, are much lighter particles than excitons
and are accelerated outwards from the center of the spot. As sketched in
Fig.~\ref{fig1}a, the cloud of expanding polaritons propagates ballistically for tens of microns, before eventually relaxing into the lowest energy level outside of the blue-shifted region.  This process of expansion and relaxation leads, above a critical density, to the formation of a condensed state at the bottom of the lower polariton branch, which extends over a wide spatial region and even at long distances from the position of the pumping spot \cite{BallariniarXiv2016}. The light emitted by the sample carries all the information about the spatio-temporal correlations of the extended polariton condensate, that can be measured by a Michelson interferometer as described in the following and in the Supplementary Information.
\begin{figure}[htbp]
\centering
	\includegraphics[width=0.48\textwidth]{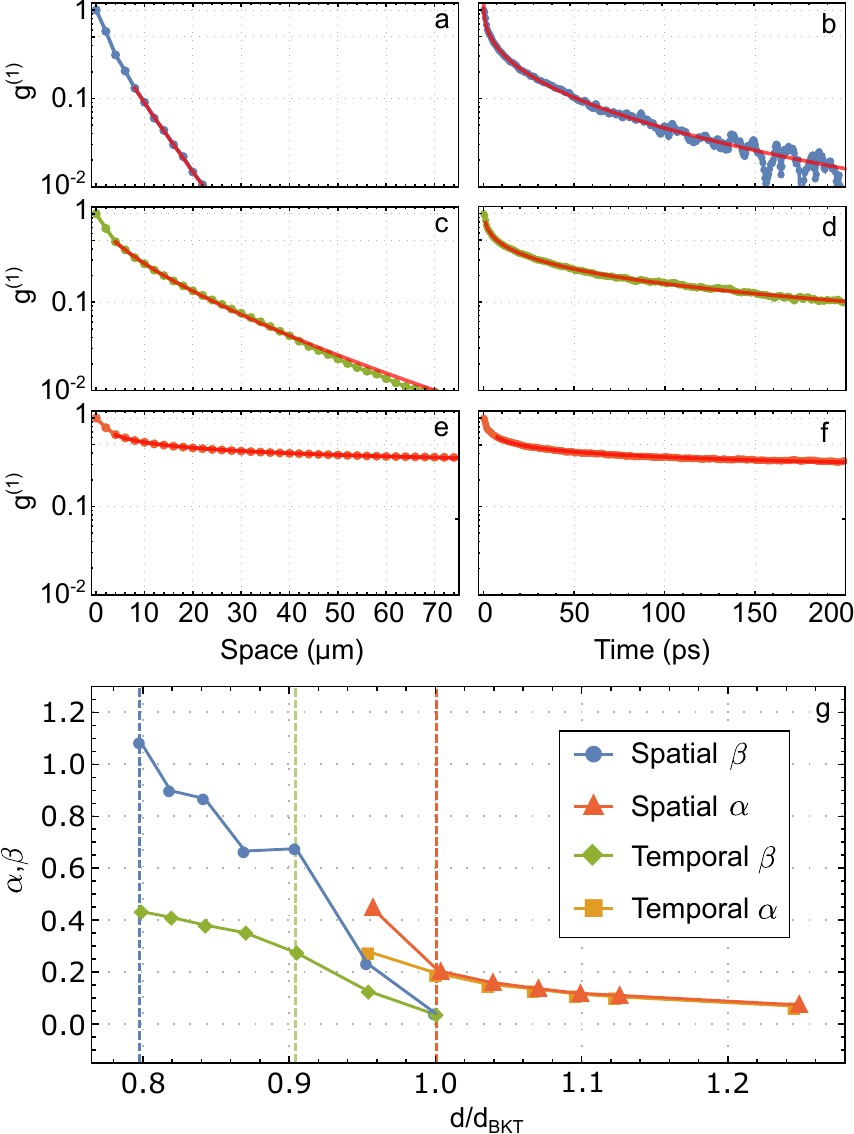}
	\caption{\textbf{Decay of coherence from stochastic
            analysis. a,c,e} Spatial decay of
          coherence. Respectively, an exponential decay, a stretched
          exponential with $\beta=0.67$ and a power law decay with
          $\alpha=0.20$. \textbf{b,d,f,} Temporal decay of
          coherence. Respectively, a stretched exponential with
          $\beta=0.41$, a stretched exponential with $\beta=0.27$ and
          a power law with $\alpha=0.20$.  These three cases are
          indicated in \textbf{g} with blue, green and red
          vertical dashed lines.
           \textbf{g,} Exponents $\beta$ of
          the stretched exponential fit, for spatial (blue) and
          temporal (green). Exponents $\alpha$ of the power law
          fitting for spatial (red) and temporal (orange).
          Note, we normalise the density $d$ to the density $d_{\rm BKT}$ 
          at which the algebraic decay clearly fits the data better 
          then the stretched exponential (see Fig.~6 in SI).
         }
\label{fig_th1}
\end{figure}
The spatial correlation is extracted from the interference pattern
obtained overlapping the 2D emission map with its centro-symmetric
image. In Fig.~\ref{fig1}b, the interference fringes are visible around the
center of the image, which is the autocorrelation point $\mathbf{r}=\mathbf{r_{0}}$. The phase coherence in space is therefore measured by the first order correlation function $g^{(1)}(\mathbf{r}-\mathbf{r_{0}})$ (where omitted, $\mathbf{r_{0}}= 0$ and $t=t'$ is assumed). In Fig.~\ref{fig2}, $g^{(1)}(\mathbf{r})$ is shown in colour-scale for different values of the polariton density. Increasing the pumping power, a higher level of coherence is sustained over larger distances (Fig.~\ref{fig2}(a-e)), up to a maximum in Fig.~\ref{fig2}e. Note that the high quality of the sample allows a uniform coherence to extend over a wide spatial region of about \SI{80}{\micro\meter}$\times$\SI{60}{\micro\meter}. Increasing further the excitation power results in the shrinking of the spatial extension of the coherence, as shown in Fig.~\ref{fig2}f.  

In Fig.~\ref{fig3}, the horizontal line profile
$|g^{(1)}(x,-x)|$ passing through $\mathbf{r_{0}}$, for $\Delta
x>0$ (with $\Delta x\equiv 2x$), is studied at increasing pumping powers
(Fig.~\ref{fig3}(a-c)). To allow a uniform
description across the transition, both a power law and a stretched
exponential functions are used in the fitting procedure:
\begin{align}
|g^{(1)}(x,-x)|&= A|2x|^{-\alpha  \label{eq01}}\\
|g^{(1)}(x,-x)|&= Ae^{-B|2x|^{\beta}} \label{eq1}
\end{align}
with $B$ a scale parameter for the $x$-axis.
At low pumping power (Fig.~\ref{fig3}a), the decay is exponential and it is well fitted by eq. (\ref{eq1}) with $\beta\approx1$. Increasing the pumping power, the exponential decay first changes into a short range envelope $\beta>1$, and then, at the threshold density $\text{d}_\text{th}$ for the nonlinear increase of the ground-state population \cite{BallariniarXiv2016}, coherence starts to build up at larger distances. This corresponds to a change in the slope of the decay, which is still best fitted by eq. (\ref{eq1}), but now with $\beta<1$ (Fig.~\ref{fig3}b). Finally, at $d\approx2.7~\text{d}_\text{th}$, a power law decay is evident (Fig.~\ref{fig3}c), with a high degree of spatial coherence (>\SI{50}{\%}) extending over distances of $\approx \SI{50}{\micro\meter}$. Remarkably, this slow decay is characterized by the exponent $\alpha=0.22$. In Fig.~\ref{fig3}g, the $\alpha$ and $\beta$ exponents are shown at different densities: while below threshold only eq. (\ref{eq1}) is able to fit the experimental data, above threshold the best fit is given by eq. (\ref{eq01}) (see Supplementary Information). 
However, as will be shown in the following, it is essential to verify that a similar behavior is also observed for the temporal correlations.

The decay of temporal coherence is extracted by using the long delay line to change the relative temporal delay of the two interferometer arms up to $200$ ps (see Methods and
Supplementary Information).
In Fig.~\ref{fig3}(d-f), the temporal coherence at the autocorrelation point $g^{(1)}(t,t+\Delta t)$ is shown for three different polariton densities.
In Fig.~\ref{fig3}h, the $\alpha$ and $\beta$ exponents of equations (\ref{eq01})
and (\ref{eq1}) that best fit the experimental data are shown across the
transition. Below threshold, coherence decays quickly and follows a Gaussian slope ($\beta\approx2$). 
At $d=1.3~\text{d}_\text{th}$, the temporal coherence can be best fitted by (\ref{eq1}) with an exponent $\beta\approx0.8$ (or, with a slightly worst fit, with a power-law of exponent $\alpha\approx0.57$), while at $d \approx 2.7~\text{d}_\text{th}$, the long time behaviour clearly follows a power law with $\alpha=0.2$. The residuals analysis proves the agreement between the experimental data and the fitting model (see Supplementary Information).  Crucially, also for time correlations, $\alpha<0.25$, which coincides, within the experimental accuracy, with the one obtained from the spatial coherence at the corresponding density.
%

We performed complementary theoretical analysis, based on the exact
solution of the stochastic equations of motions \cite{Galaa2015} (see
Methods), with the same microscopic parameters as the ones of
the experiment. We observed the same crossover from an exponential via stretched exponential to an algebraic decay of coherence in space and time (Fig.~\ref{fig_th1}). In particular, for our long polariton lifetime, we see the spatial and temporal $\alpha$
being the same and always smaller then 1/4 above the BKT threshold 
(Fig.~\ref{fig_th1}g), showing that the drive and dissipation do not prevail in this good quality sample in
contrast to the earlier studied non-equilibrium cases \cite{Galaa2015,yamamoto}. 
\begin{figure}[htbp]
\centering
	\includegraphics[width=0.48\textwidth]{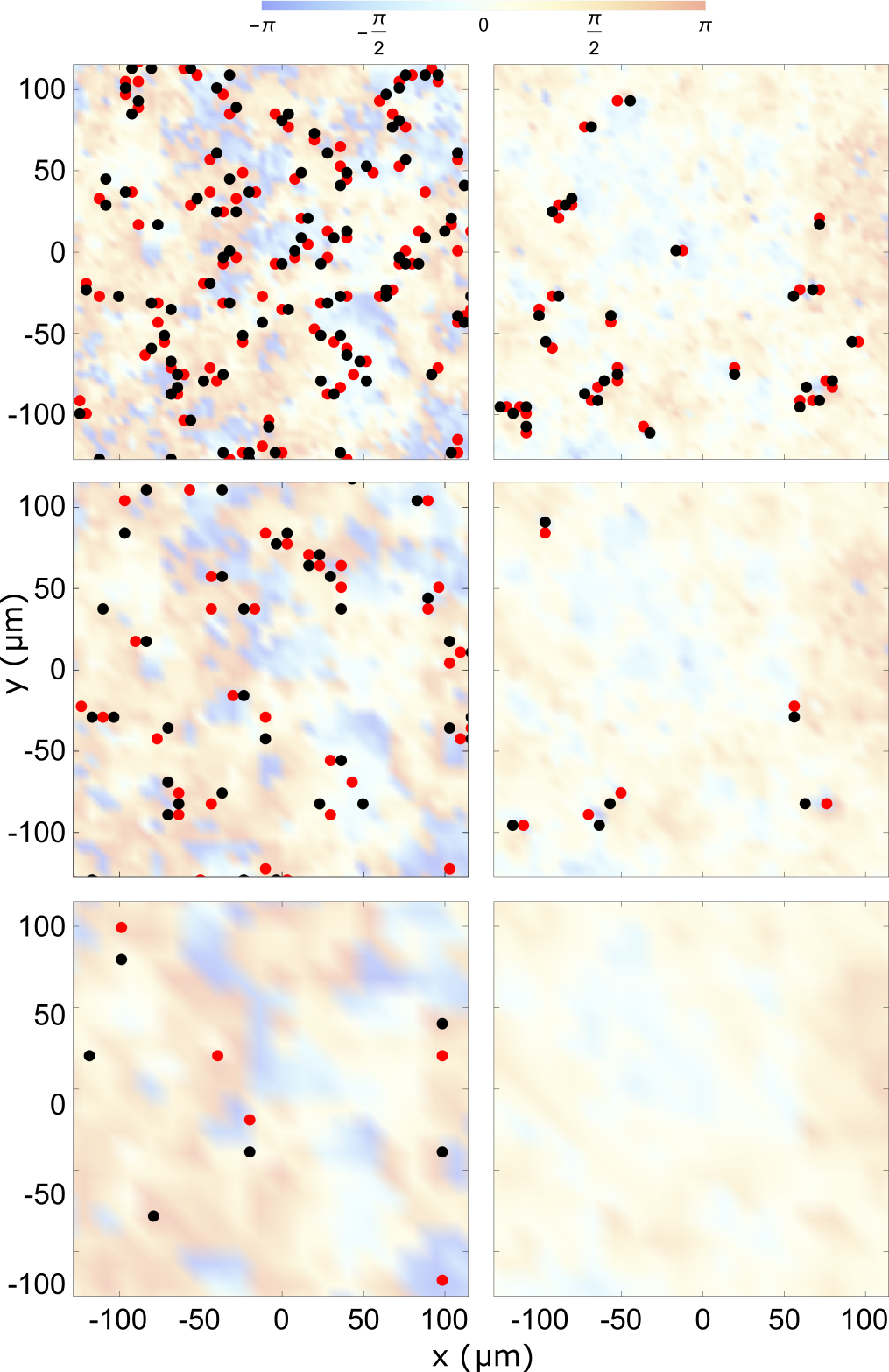}
	\caption{\textbf{Vortex-antivortex distribution map.} 
	{\textbf{ Top,}} Vortices (V) in red and anti-vortices (AV) in
          black just before (left) and after (right) the BKT transition
          with parameters as in Fig.~\ref{fig_th1}c and e, respectively.
          \textbf{Middle-Bottom} The same as in \textbf{Top} but after
          filtering off in two steps high momentum states to eliminate bound
          pairs. Such filtering reveals the presence of free
          vortices. Note that there are no
          free vortices when spatial and temporal coherence show
          algebraic decay (right) but there are some free vortices
          in the case of stretched exponential decay of coherence (left).
          The underlying colour map shows the phase profile of the
          field.  }
\label{fig_th3}
\end{figure}
Additionally, while the vortex-antivortex binding cannot be directly
observed in the experiments, which average over many realizations, the numerical analysis is able to track the presence of free vortices in each single realisation, confirming the topological origin of the
transition. Indeed, we see clearly that, in the algebraically ordered
state, free vortices do not survive and the pairing is complete
(Fig.~\ref{fig_th3} right). In contrast, the exponential and stretched-exponential regimes both show the presence of free vortices (Fig.~\ref{fig_th3} left), 
the number of which decreases as we move across the transition.
Since the stretched exponential phase is always associated with some 
presence of free vortices, this supports 
that we are observing a BKT crossover rather than a Kardar-Parisi-Zhang phase
\cite{PhysRevX.5.011017}.

\begin{figure}[htbp]
\centering
	\includegraphics[width=0.48\textwidth]{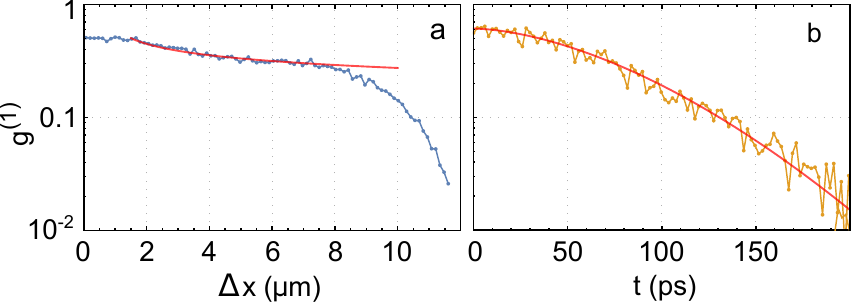}
	\caption{\textbf{Spatial and temporal coherence in weak
            coupling regime.} \textbf{a,} Spatial coherence showing a power law
          decay with $\alpha=0.253$.
          \textbf{b,} Temporal decay of coherence with stretched
          exponential fitting exponent $\beta=1.8$.}
\label{fig5}
\end{figure}

Finally, to highlight the importance of studying correlation not only in the space but also in the temporal domain, we used a microcavity in the weak-coupling regime, showing that a power law decay of coherence in space is associated in this case to a Gaussian
decay in time. Using a sample with a lower quality factor and less quantum wells, we induce, under high non-resonant pumping, the photon-laser regime (as in a VCSEL) \cite{butte2002,Bajoni2008b}. Despite the fact that this system is clearly out-of-equilibrium, it shows a power law decay of spatial coherence with $\alpha= 0.25$ (Fig.~\ref{fig5}a) within the pumping spot region (with a radius of about $\SI{10}{\micro \meter}$). Remarkably, the behavior of spatial correlations is very similar to what obtained in Ref\citen{yamamoto}, but the temporal coherence, shown in Fig.~\ref{fig5}b, follows a quasi-Gaussian decay (stretched exponential with $\beta$ exponent about 1.8), not compatible with the algebraic order characteristic of the BKT phase. This shows that a consistent behavior between time and space is necessary to evidence the BKT transition in driven/dissipative systems.

\section*{Conclusions}

In conclusion, we observed the formation of an ordered phase in two-dimensional driven/dissipative ensemble of bosonic quasiparticles, exciton-polaritons in semiconductor microcavities, exploring both spatial and temporal correlations across the transition. This system lies at the interface between equilibrium and out-of-equilibrium phase transitions, and it has been often compared both to Bose-Einstein condensates and to photon lasers. We show that the measurement of spatial correlations $g^{(1)}(\mathbf{r})$ alone is not sufficient to establish whether an open/dissipative system is at equilibrium. Instead, two distinct measurements, one in time and one in space domain, are needed. Satisfying this requirement, we report a power-law decay of coherence with the onset of the algebraic order at the same relative density and comparable exponents for both space and time correlations. 
We should stress that the exceptionally long polariton lifetime in the present sample allows us to reach the BKT phase transition at lower densities, and in the region without the excitonic reservoir, resulting in a lower level of dephasing. In our experiments, the absence of any trapping mechanism, be it from the exciton reservoir or potential minima, allows us to avoid the influence of finite-size effects in the temporal dynamics of the autocorrelation \cite{PhysRevB.75.195331}. 
Simulations with stochastic equations match perfectly the experimental results and demonstrate that the underlying mechanism of the transition is of the BKT type i.e. a topological ordering of free vortices into bound pairs, resulting in the coherence built up  both in space and time. All these observations validate that polaritons undergo phase-transitions following the standard BKT picture, and fulfill the expected conditions of thermal equilibrium despite their driven/dissipative nature. Now that the equilibrium character of polaritons becomes a tunable parameter, the study of driven/dissipative phase transitions and of the universal scaling laws is within reach in this solid state device.

\section*{Methods}

In the present experiment, polaritons were created in a planar
semiconductor microcavity. The sample is described in detail in Ref.\citen{BallariniarXiv2016} and in the Supplementary Information.
Polaritons are injected in the system by making use of a single-mode Ti:sapphire ring laser in continuous wave (CW) operation and tuned at energy around $\SI{1686.80}{\milli \eV}$ with a stabilized output wavelength and power (M2Squared SolsTis) to reduce fluctuations in the exciton reservoir. In particular, the 
energy of the laser is chosen to coincide with a minimum of the reflection stop band. To avoid sample thermal heating the pump laser is chopped at a frequency of $4$ kHz with a duty cycle of $8\%$. The chopper window aperture time (in the scale of \si{\micro\second}) is significantly larger than the polariton lifetime so it does not affect the dynamics. The excitation laser beam is focused in a Gaussian spot with a FWHM $\approx$ \SI{20}{\micro\meter}. The Gaussian laser profile in space is the best one to trigger the expulsion-relaxation mechanism, allowing the creation of a sufficiently uniform and extended polariton condensate.
On the detection line, the emission from the sample is collected in epilayer configuration and it is sent to a Michelson interferometer with two delay lines as showed in Fig.~S1 of SI.
On the first interferometer arm, a short (fraction of the wavelength) piezo stage motorized is used to move a mirror. This allows the reconstruction of the spatial coherence $g^{(1)}(\mathbf{ r} - \mathbf{r_0})$ using the sinusoidal envelope of the intensity (see
Supplementary Fig.~S2).
On the second arm, a long (about $200$ ps) motorized screw actuator is used to change the position of the back-reflector to characterize the behavior of the coherence in time. By using the back-reflector, one of the two arms is centrosymmetrically rotated respect to the other and these two beams interfere on the focal plane of a CCD (Charge-Coupled Device) camera. To obtain energy resolved spatial and momentum emission images, a spectrometer is placed before the CCD detector.

\section*{Acknowledgements}
Funding from the POLAFLOW ERC Starting Grant is acknowledged.
M. H. S. acknowledges support from EPSRC (Grants No. EP/I028900/2 and No. EP/K003623/2).

\section*{Author contributions}
D.C. and D.B. took and analysed the data. G.D. and M.H.S. performed stochastical numerical simulations. C.S.M. and F.P.L. discussed the results. D.C.,D.B, C.S.M., M.D.G., L.D., G.G, F.P.L. M.H.S. and D.S. co-wrote the manuscript. K.W. and L.N.P. fabricated the sample. D.S. coordinated and supervised all the work.


\newpage

\pagebreak
\widetext
\begin{center}
\textbf{\large Topological order and equilibrium in a condensate of exciton-polaritons\\
Supplemental Material}
\end{center}
\setcounter{equation}{0}
\setcounter{figure}{0}
\setcounter{table}{0}
\setcounter{page}{1}
\makeatletter
\renewcommand{\theequation}{S\arabic{equation}}
\renewcommand{\thefigure}{S\arabic{figure}}
\renewcommand{\bibnumfmt}[1]{[S#1]}
\renewcommand{\citenumfont}[1]{S#1}

\section*{Sample and experimental setup}

The sample used in this work is a high quality factor (Q>100000) $\rm GaAs$/$\rm AlGaAs$ planar microcavity containing 12 $\rm GaAs$ quantum wells of 7 nm width, grouped in 3 blocks placed at the antinode positions of the electric field inside the cavity, with a collective Rabi splitting of $\hbar\Omega=\SI{16}{\milli\electronvolt}$. The front (back) mirror consists of 34 (40) pairs of $\rm AlAs$/$\rm Al_{0.2}Ga_{0.8} As$ layers, with a polariton lifetime of about 100 ps \cite{BallariniarXiv2016S}.
The cavity detuning $\delta=\hbar(\rm \omega_{c}-\rm \omega_{x})$ is slightly negative ($\delta=\SI{-1}{\milli\electronvolt}$), with the exciton energy $\rm \hbar \omega_{x}$=\SI{1611.2}{\milli\electronvolt}.
The sample is placed in a cryostat and kept at a temperature of about 10 K.

\begin{figure}[htbp]
\centering
	\includegraphics[width=0.48\textwidth]{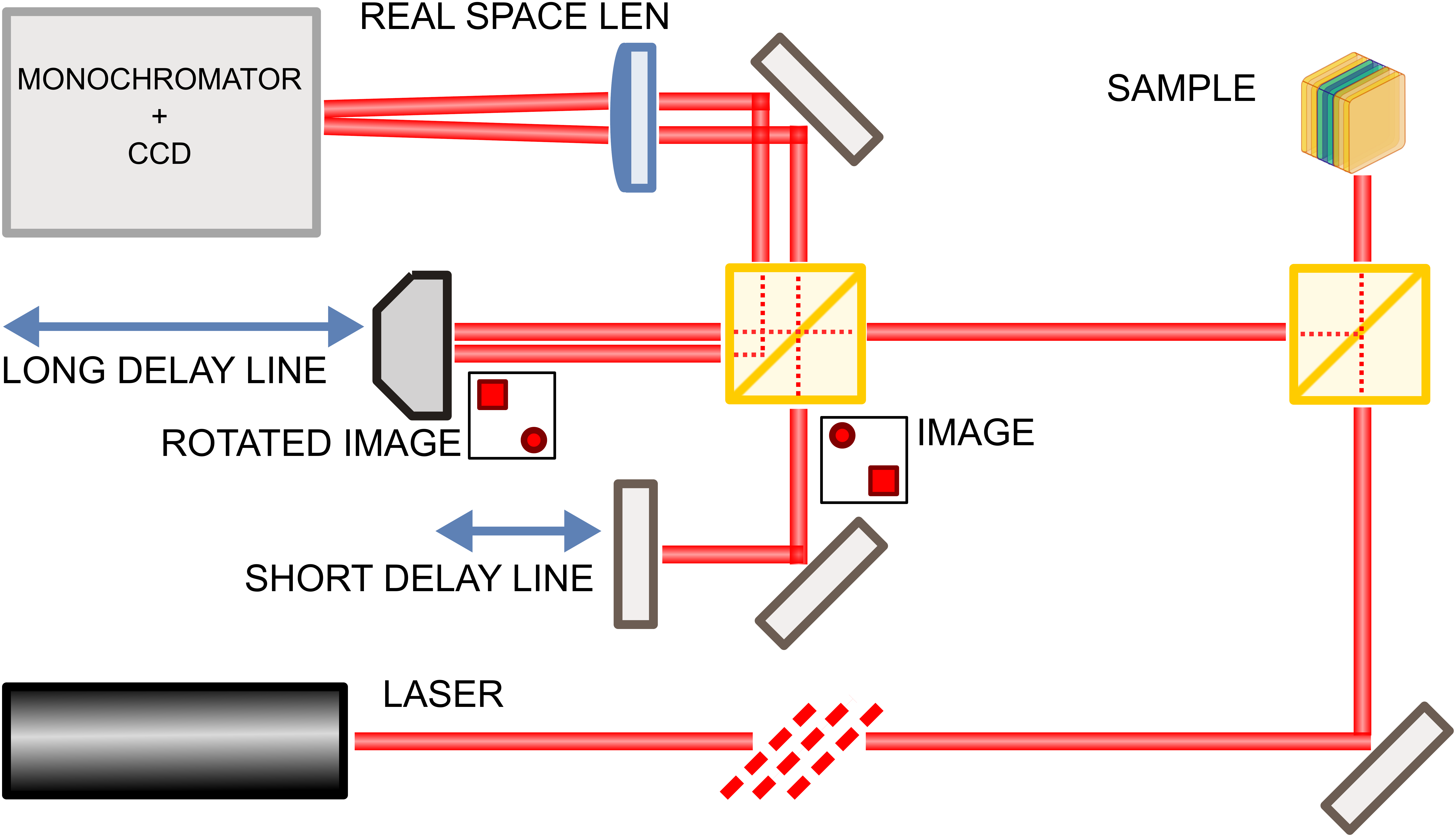}
	\caption{Michelson interferometer with two different delay lines: a first short one to measure the spatial decay of the correlation and a long one to evaluate the temporal decay.
 }
\label{setup}
\end{figure}

The coherence properties are investigated using a Michelson interferometer with two different delay lines, as showed in Fig.~\ref{setup}.

\section*{First-order Correlation Function Decay}

\begin{figure}[htbp]
\centering
	\includegraphics[width=0.48\textwidth]{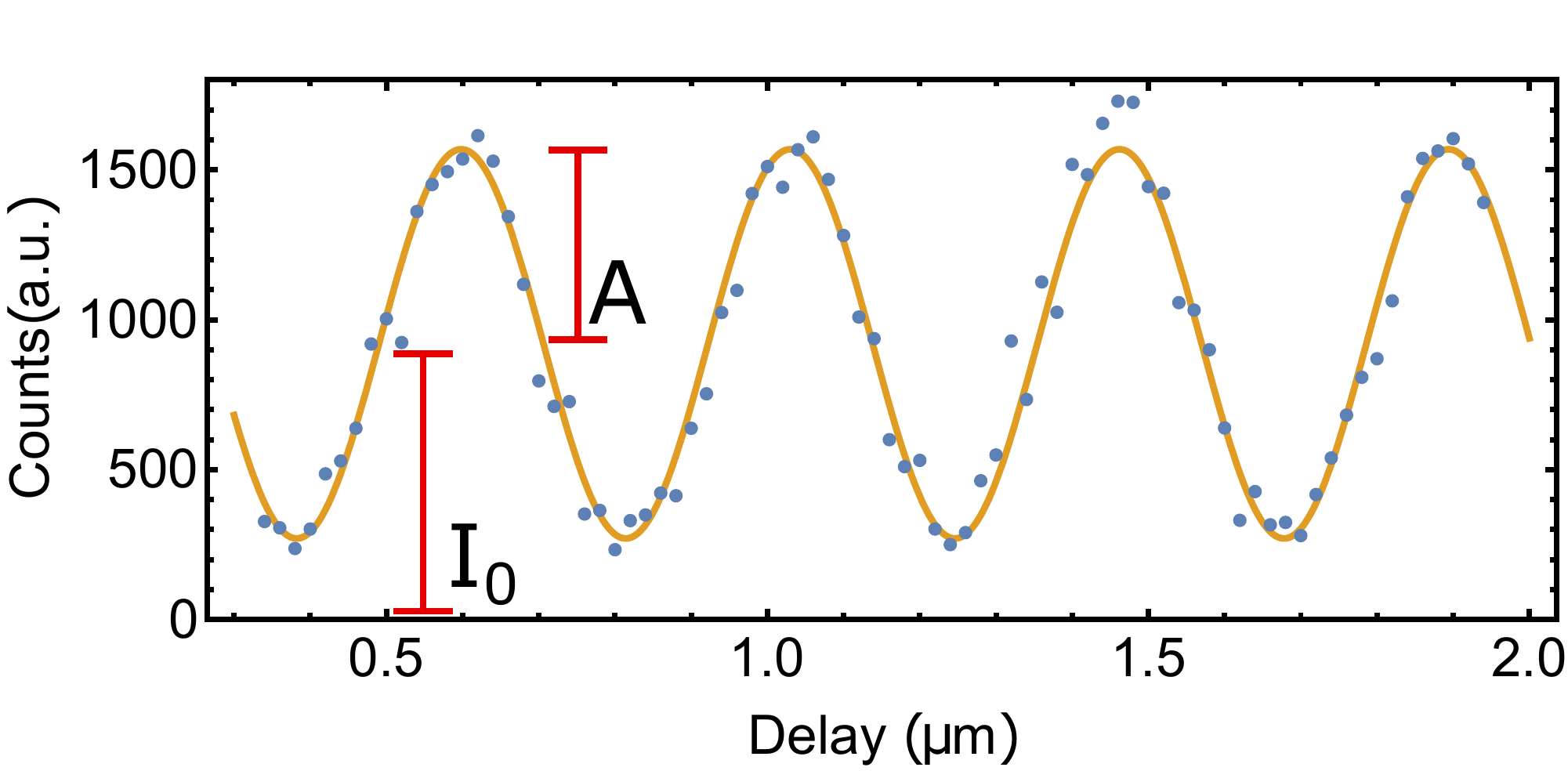}
	\caption{Sinusoidal envelope of the intensity on a single CCD pixel as a function of the piezo-controlled delay between the two arms of the interferometer.
 }
\label{fig:fig6}
\end{figure}

The first order correlation function is defined as:
\begin{equation}
	g^{(1)}(r_1,t_1;r_2,t_2)=\frac{\langle \psi^{*}_1 \psi_2
          \rangle}{\sqrt{\langle \psi^{*}_1 \psi_1 \rangle \langle
            \psi^{*}_2 \psi_2 \rangle}}
\label{eq:g1}
\end{equation}
with $\psi^{*}_i$ and $\psi_i$ the creation and annihilation operators for the space-time point $(r_i,t_i)$, with $i=1,2$.
By changing the length of one path through the use of a motorized piezo stage, the relative phase is changed, and the intensity of each point of the image shows a sinusoidal envelope as shown in Fig.~\ref{fig:fig6}.
The path length is scanned by sub-wavelength steps in order to extract the first order correlation function as: 
\begin{equation}
|g^{(1)}(\mathbf{r},\mathbf{-r})|= V I_{\text{ideal}},
\end{equation}
where $I_\text{ideal}=\displaystyle (I_1+I_2)(2\sqrt{I_1I_2})^{-1}$
takes into account small asymmetries between the two interferometer arms, with $I_1$ and $I_2$ the intensities in the two interferometer arms and $V$ is the visibility of the interference fringes obtained  by fitting the data with:
\begin{equation}
	I(x)=I_0 + A \sin(\omega x + \phi_0)
\end{equation}
where the visibility $V=\frac{A}{I_0}$ is shown in Fig.~\ref{fig:fig6} and $\phi_0$ is the initial phase.

\section*{Coherence Length and Time}

In order to evaluate the coherence length and time as a function of density, both spatial $g^{(1)}(x,-x,t=0)$ and temporal $g^{(1)}(t, t+\Delta t)$ are fitted with stretched exponentials as reported in Eq.~(2) of the main text. It is therefore possible to define both in space and time the relaxation length (time) as:
\begin{equation}
\displaystyle
\langle l \rangle= \int_0^\infty dx e^{-(x/l_e)^{\beta}} = \frac{l_e}{\beta} \Gamma \left(\frac{1}{\beta} \right)
\end{equation}
with $l_e$ the renormalization factor scale of the x-axis and $\Gamma$ the gamma function, with x-axis both the spatial and temporal one. The resulting coherence length and time are reported in Fig.~\ref{fig:fig7}:
\begin{figure}[htbp]
\centering
	\includegraphics[width=0.45\textwidth]{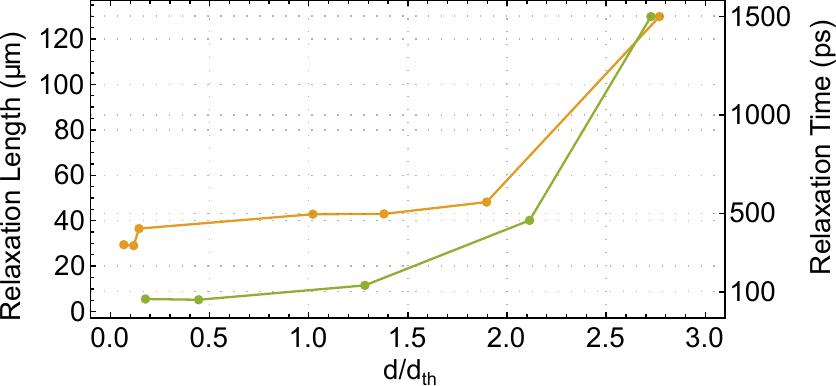}
	\caption{Coherence length (yellow) and time (green) for different densities extracted from the stretched exponential fitting.
 }
\label{fig:fig7}
\end{figure}

\section*{Fitting Model and Residuals Analysis}

\begin{figure}[h]
\centering
	\includegraphics[width=0.5\textwidth]{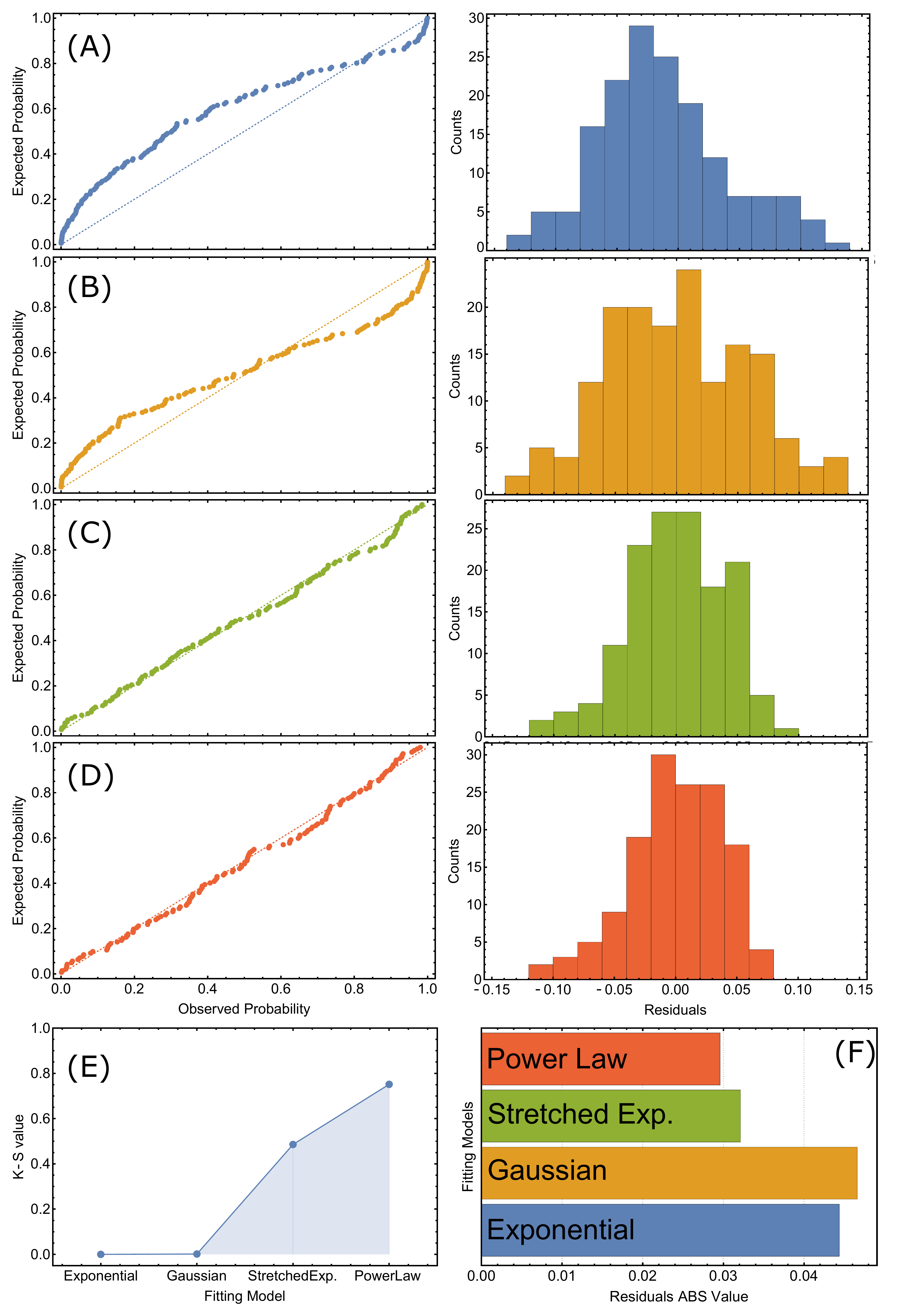}
	\caption{Fitting residuals analysis for the temporal algebraic regime (red square in Fig.~3h of the main text). Each color corresponds to a different model used to fit the temporal decay. Row \textbf{a}: P-P plot for the observed and expected Gaussian with minimum standard deviation of residuals distribution for the exponential model.
    Row b: Gaussian model.
    Row c: Stretched exponential model.
    Row d: Power law model.
    e: Kolmogorov-Smirnov value quantifying the normality of the residuals distribution for each model.
    f: Total Absolute Value of fitting residuals, normalised to the number of point used.
 }
\label{fig:fig8}
\end{figure}

\begin{figure}[h]
\centering
\includegraphics[width=0.5\textwidth]{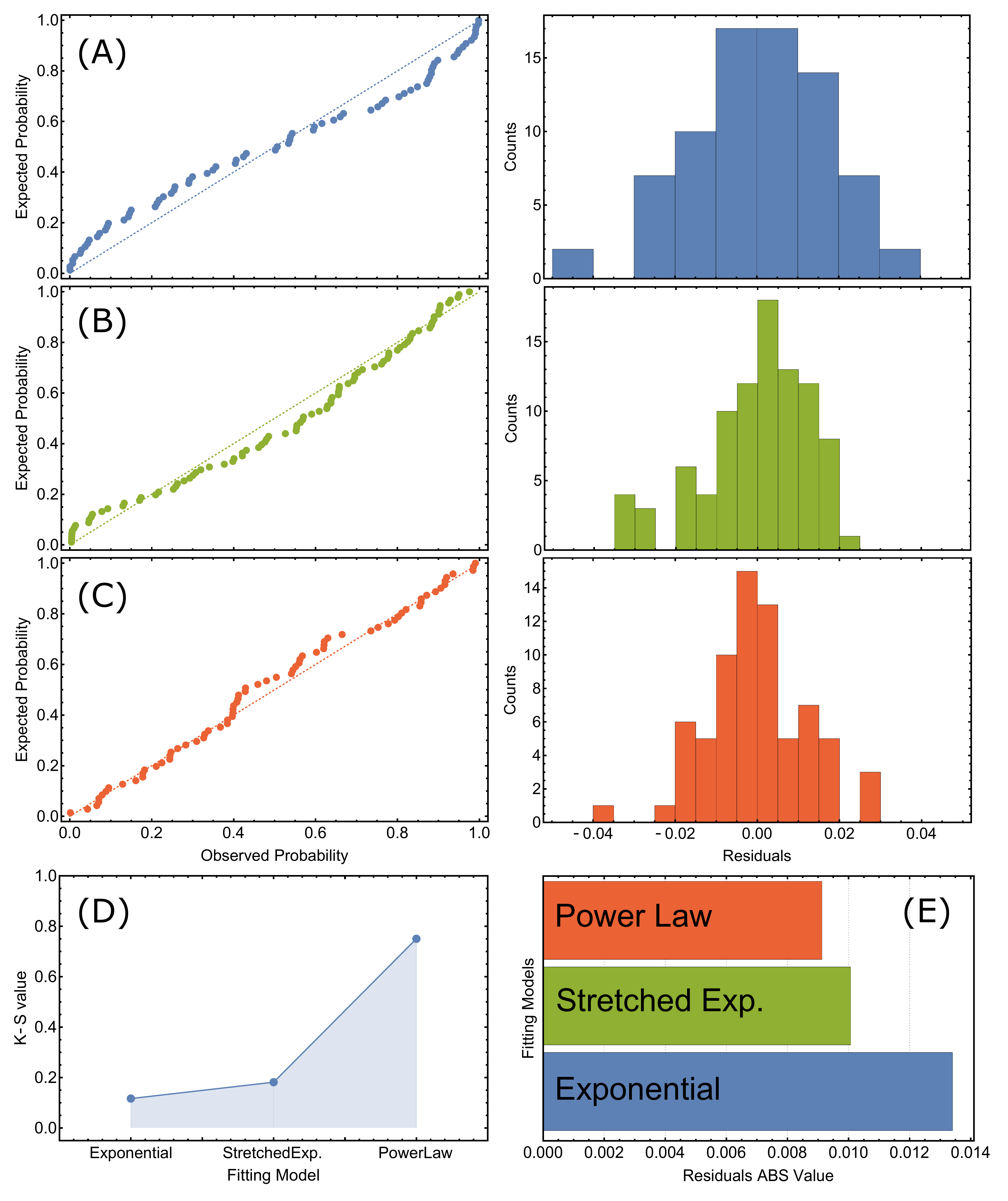}
\caption{Fitting residuals analysis for the spatial algebraic regime (red square in Fig.~3g of the main text). Each color corresponds to a different model used to fit the spatial decay. Row a: P-P plot for the observed and expected Gaussian with minimum standard deviation of residuals distribution for the exponential model.
Row b: Stretched exponential model.
Row c: Ppower law model.
d: Kolmogorov-Smirnov value quantifying the normality of the residuals distribution for each model.
e: Total Absolute Value of fitting residuals, normalised to the number of point used.
} 
\label{fig:fig9}
\end{figure}

The analysis of the fitting results allows to assess the applicability  of the model used in the fitting procedure. In Fig.~\ref{fig:fig8} we show the residuals of fitting the temporal first order correlation function $g^{(1)}(t, t+\Delta t)$ in the quasi-ordered regime (Red square in Fig.~3h of the main paper). We report different attempts to fit the data using an exponential (blue, row (a)), a Gaussian (yellow, row (b)), a stretched exponential (green, row (c)) and a power law (red, row (d)). The P-P (Probability Probability) plot is a standard tool to investigate the deviance of a data set from the normal distribution.
Indeed, fitting residuals with a normal distribution around the zero value represents a strong indication that the power-law model used to fit the data is the best one. The Kolmogorov-Smirnov test permits to quantitatively check the deviation from the normal distribution of a dataset. In order to assure that we have the minimum residual spreading, we used a normalised Gaussian distribution with $\sigma$ the minimum from the residuals distributions for each model fitting the experimental data. The value of this test, maximum for the power law model, combined with the fact (Fig.~\ref{fig:fig8}f) that the sum of the absolute values of the residuals is minimum for the power law model, confirms that the power law is the best fitting for the temporal decay of the correlation above BKT transition. In Fig.~\ref{fig:fig9} we show the  same analysis for the same regime (red square in Fig.~3g of the main paper but for the spatial decay of correlations. In this case the Gaussian fitting is not shown because the values of the residuals is too large.

\section*{Theoretical Description}

In the classic equilibrium BKT scenario, for a system with linear
dispersion in the ordered phase, we expect a slow algebraic decay (as
eq. (1) of the main text) of the first order coherence with exactly
the same power-law exponents in {\it both} space ($\alpha_{s}$) and
time ($\alpha_{t}$) \cite{Szymanska2006S,PhysRevB.75.195331S}, given by
$\alpha_{s,t}=k_BT/n_s$, where $T$ indicates temperature and $n_s$ the
superfluid density.  $\alpha_{s,t}$ have an upper bound of 1/4
\cite{Nelson1977S} at which density/temperature vortices proliferate,
causing an exponential decay of coherence, characteristic for the
disordered phase.  At the same time non-equilibrium dissipative driven
systems, with diffusive spectrum in the ordered phase
\cite{Szymanska2006S,Wouters2007}, have been shown to still exhibit an
algebraic decay of coherence {\it but} with temporal correlations
decaying two times slower then the spatial ones $\alpha_{t}=1/2
\alpha_{s}$ \cite{Szymanska2006S,PhysRevB.75.195331S}. Moreover, values
of $\alpha_{s}$ as large as four times the equilibrium upper bound,
when approaching the BKT transition, were reported both experimentally
\cite{yamamotoS} and from theoretical analysis \cite{Galaa2015}, using
beyond-mean-field truncated Wigner methods able to account for
vortices, suggesting an ``over-shaken'' superfluid state
\cite{Galaa2015S}.
Finally, it has recently been suggested that the dissipation might in
fact have an even more profound effect on the system with collective
phase fluctuations destroying the algebraic order at long distances,
leading to a stretched exponential decay of first order coherence
characteristic of Kardar-Parisi-Zhang phase
(KPZ)\cite{PhysRevX.5.011017S}. In that scenario the parameter $\beta$
of the stretched exponential (see Eqn. (2) of the main paper) are also
different for space ($\beta \approx$ 0.78) and time ($\beta \approx $
0.48). Even if later estimates of the KPZ length-scales appeared to be
unrealistic, and the presence of free topological defects strongly
hampers the possibility of the KPZ phase \cite{Altman2016S}, the true
nature of the 2D exciton-polariton phase transition and the resulting
order is still at the center of an intense debate. Additionally, the
type of the Renormalisation-Group (RG) analysis, which led to those
conclusions \cite{PhysRevX.5.011017S,Altman2016S}, rely on the expansion
in one over the mean-field density, and so are inadequate in
describing the crossover region close to the phase transition. Thus,
here, to specifically address the region close to the phase transition
we restore to exact numerical solutions of the stochastic equations of
motions, described in the next section.

\section*{Stochastic simulations}

Our system consists of an ensemble of bosonic particles (the
lower-polaritons) of mass $m$ and lifetime $2\kappa$, interacting via
contact interactions $g$, and driven incoherently with pump of strength
$\gamma$. The pumping saturation due to other processes is 
$\Gamma$. Using Keldysh field theory one can show that by including
the classical fluctuations to all orders, but quantum fluctuations only
to the second order (correct in the long-wavelength limit), and employing
the MSR formalism, one can arrive at the stochastic equation for
the field $\psi(\mathbf{r},t)$ (for review see \cite{Diehl_review}).
Alternative derivation, using the Focker-Plank equation for the
Wigner function, aimed at numerical implementation on a finite
spatial grid $dV $has been reviewed in \cite{Carusotto2012}. The finite grid
version reads
\begin{equation}
\label{eq:sgpe}
 i d\psi(\mathbf{r},t) = 
  \biggl [-\frac{\nabla^2}{2m} +  g |\psi(\mathbf{r},t)|_{-}^2 + (\gamma -
       \kappa -\Gamma  |\psi(\mathbf{r},t)|_{-}^2 
       )   \biggr]\psi(\mathbf{r},t)
 dt+ dW
\end{equation}
where $dW$ is the Wiener noise with correlations 
\begin{equation*}
\left\langle
  dW^*(\mathbf{r'},t)dW(\mathbf{r},t)\right\rangle=
\frac{\gamma + \kappa + \Gamma
     |\psi(\mathbf{r},t)|_{-}^2}{dV}
\delta_{\mathbf{r},\mathbf{r'}}dt, 
\end{equation*}
where by $|\psi(\mathbf{r},t)|_{-}$ we abbreviated the following
expression for the density
$|\psi(\mathbf{r},t)|_{-}=|\psi(\mathbf{r},t)|-\frac{1}{dV}$, which
comes from the Wigner function relating to the time symmetric and not
time ordered operators.
In Ref. \cite{PhysRevX.5.011017S} the Eqn. (\ref{eq:sgpe}) is solved
approximately analytically using RG analysis by eliminating the
amplitude fluctuations, and focusing solely on the non-topological
phase-fluctuations. The procedure, however, relies on an expansion in
one over the mean-field density, which together with approximating the
amplitude fluctuations and discarding vortices, cannot capture the
region close to the phase transition. Instead, here, we solve this
equation exactly numerically on a finite grid. In addition, we have
checked our results with a more general model of a saturable drive,
applicable up to higher densities, in which $\gamma -\Gamma
|\psi(\mathbf{r},t)|_{-}^2$ in Eqn. (\ref{eq:sgpe}) is replaced by
$\frac{\gamma}{1+\frac{|\psi(\mathbf{r},t)|_-^2}{n_s}}$ with the noise
strength proportional to
$\frac{\gamma}{1+\frac{|\psi(\mathbf{r},t)|^2_-}{n_s}} +
\kappa$. There is no appreciable difference between the results from
the two models for our relatively low densities.

We evolve the dynamics of the stochastic equations~\eqref{eq:sgpe}
with the XMDS2 software \cite{Dennis2012} using a fixed-step (to
ensure stochastic noise consistency) 4th order Runge-Kutta (RK)
algorithm, which we have tested against fixed-step 9th order RK, and a
semi-implicit fixed-step algorithm with 3 and 5 iterations.
We choose the system parameters the same as in the experimental part:
the mass of the microcavity lower polaritons is taken to be
$m_C=3.8\times10^{-5}m_e$, where $m_e$ is the electron mass, the
polariton lifetime $2\kappa =101.3ps$, and the polariton-polariton
interaction strength $g=0.004$~meV$\mu$m$^2$. The only parameter not
possible to extract from the experiment is the saturation rate of the
driving process (or in other words the three-body type losses in the
system). We perform analysis for a range of $\Gamma$ (or $n_s$)
values, and since the other parameters are fixed, we choose the value
of $\Gamma$ ($n_s$)to reproduce the overall length scale of $g^{(1)}(x)$.

In our method, the stochastic averages over the configurations of
different realisations of the fields provide the expectation value of
the corresponding symmetrically ordered operators, and it is important
to get the results to converge in the number of realisations.  The
first order spatial correlation function $g^{(1)}(x)$ is evaluated
according to Eq.~\eqref{eq:g1}, by averaging over 100 independent
stochastic paths, and additionally over auxiliary position in space
$r_0$, since in simulations our system is uniform. The temporal
correlation function is evaluated from a single spatial point (to
avoid picking up any spatial correlations) after the steady-state is
reached, and averaged over 10000 stochastic paths. Since the Wigner
average provides the expectation value of the corresponding {\it
symmetrically} (and not time)-ordered operators, we need to subtract
the expectation value of the commutator. For single time correlation
functions, such as $g^{(1)}(x)$, the commutator is simply
$\frac{1}{2dV}$. For two-time correlation functions, such as
$g^{(1)}(t)$, strictly speaking the expectation value of the
commutator is unknown. It is, however, changing from $\frac{1}{2dV}$
at $t=0$ to $0$ at $t\to \infty$. Using the two limiting values allows
us to estimate the error, which we expect to be small given the
densities considered. Indeed, for the densities used here the
difference is practically indistinguishable.
 \begin{figure}[htbp]
\centering
	\includegraphics[width=0.48\textwidth]{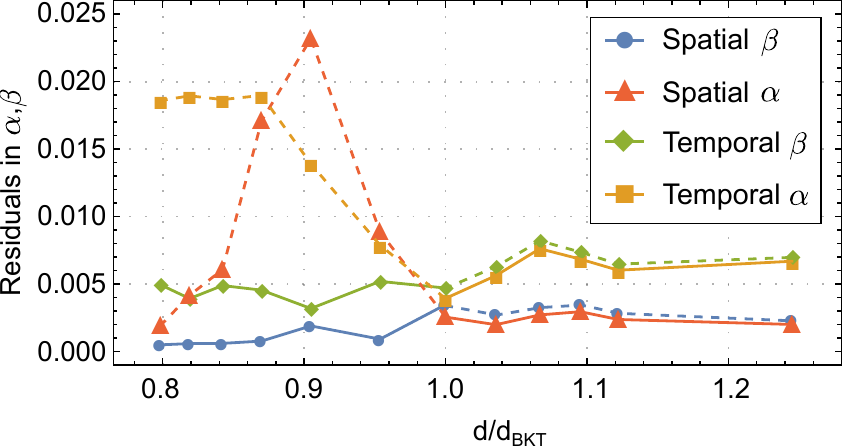}
	\caption{Residuals of the stretched exponential fit, $\beta$,
          for spatial (blue) and temporal (green) $g^{(1)}$, and the
          power-law fit, $\alpha$, for spatial (red) and temporal
          (orange) $g^{(1)}$ as a function of the polariton density
          normalised to the BKT threshold.         }
\label{fig_th4}
\end{figure}
In order to test the robustness of our conclusions to the choice of
the numerical parameters, we perform simulations on different spatial
grids varying from 2 to $\SI{5}{\micro m}$ in the grid spacing, and different
system sizes from 256 to $\SI{1024}{\micro m}$. Note, that to satisfy the
condition necessary to derive the discrete version of
equation~\eqref{eq:sgpe}, $g/[(\kappa+\gamma) d V] \ll 1$, whilst
maintaining a sufficient spatial resolution and, at the same time, a
large enough momentum range, the window of available momentum grids is
quite narrow. We find the main conclusions of our work (such as the
crossover in $g^{(1)}(x)$ and $g^{(1)}(t)$ from exponential to power-law,
spatial and temporal $\alpha$ being the same in the algebraic phase
and smaller then 1/4, as well as the behaviour of vortices)
independent on the choice of those parameters. Here, we
present a case with $\SI{2}{\micro m}$ grid spacing, and system size of
$\SI{256}{\micro m}$, which is sufficiently larger from experimental to avoid any boundary
effects influencing the relevant region.
Finally, in order to assess which functional form fits the numerical
data best we perform residuals analysis similar to those applied to
experimental data. The residuals for the data and fittings presented in the main text is shown in Fig~\ref{fig_th4}.

\end{document}